\documentclass[dvips]{acta}
\usepackage{supertabular,lscape,epsfig}
\usepackage{amssymb}
\usepackage{amsmath}

\DeclareSymbolFont{ppa}{OT1}{ppl}{m}{it}
\DeclareMathSymbol{\vv}{\mathalpha}{ppa}{'166}

\begin{document}

\renewcommand{\and}{{\rm and }}

\begin{Titlepage}
\Title{On Estimating Non-uniform Density Distributions using N Nearest Neighbors}
\Author{P.~R.~~W~o~\'z~n~i~a~k$^1$,~~and~~A.~~K~r~u~s~z~e~w~s~k~i$^2$}
{$^1$Los Alamos National Laboratory, Mail Stop B244, Los Alamos, NM~87545, USA\\
e-mail: wozniak@lanl.gov\\
$^2$Warsaw University Observatory, Al.~Ujazdowskie~4, 00-478~Warszawa, Poland\\
e-mail: ak@astrouw.edu.pl}
\end{Titlepage}

\Abstract{We consider density estimators based on the nearest neighbors method
applied to discrete point distibutions in spaces of arbitrary dimensionality.
If the density is constant, the volume of a hypersphere centered at a random location
is proportional to the expected number of points falling within the hypersphere radius.
The distance to the $N$-th nearest neighbor alone is then a sufficient statistic for the density.
In the non-uniform case the proportionality is distorted. We model this distortion by normalizing
hypersphere volumes to the largest one and expressing the resulting distribution in terms of
the Legendre polynomials. Using Monte Carlo simulations we show that this approach can be used
to effectively address the tradeoff between smoothing bias and estimator variance for sparsely sampled
distributions.}{methods: statistical, methods: numerical} 

\Section{Introduction}

Calculations based on proximity relations with nearest neighbors appear
in a wide variety of astronomical problems. The distance to $N$-th nearest neighbor
can be converted into a measure of density by means of simple inversion. Pioneering uses
of this technique in astronomy include von Hoerner (1963) and Dressler (1980).
It is known (Casertano and Hut 1985) that such conversion biases density estimates
by a factor of $N/(N-1)$ and increases their variance by $N/(N-2)$, where $N$ is the number
of considered nearest points. Density estimators based on $N=1$ or $N=2$ are therefore of little use
and at $N=4$ half of the available information is lost. Details of procedures applied in practice
and the relative merits of various choices of $N$ are reviewed by Schmeja (2011), Haas \etal (2012) and
Muldrew \etal (2012). It turns out that the adopted value of $N$ is typically between
3 and 10. The most frequently used values are 3, 4, and 5, where the effect of diminishing accuracy is large.
However, the investigators have a good reason to keep $N$ small. When the density varies in space, its
estimate based on the nearest neighbors method is effectively an average over the volume set by the $N$-th
nearest neighbor and differs from the local value. This smoothing bias is unavoidable in the case of 
variable density and independent of the rarely mentioned reciprocity bias described by Casertano and Hut (1985).
Choosing small $N$ limits the influence of the smoothing bias for the price of increasing the variance.
Another way of diminishing the smoothing bias was introduced by Ivezi\'c \etal (2005) and Cowan and Ivezi\'c (2008)
who take a ``Bayesian'' approach to combine contributions from all $N$ nearest neighbors. The net effect is, again,
lower bias at the cost of increased variance. Here we propose a new method of dealing with the smoothing bias
that captures the information on density variations contained in distances to all $N$ nearest neighbors
using the Legendre series expansion.

\Section{Uniform Density Distribution}

In this section we rederive the formula for the mean density in the case of a uniform point distribution.
In our derivation we particularly emphasize an alternative, and in fact more natural, approach to the
problem that adopts the volume per point instead of the mean density of points as the basic unknown.
This also serves as an introduction to the non-uniform case described in the next section.

Let us consider a metric space with an arbitrary number of dimensions. We will assume
that the space is populated by randomly distributed pointlike objects in such a way that
the expectation value of the number of objects $n$ contained in an arbitrarily chosen subspace
of volume $v$ is proportional to the volume $v$ with a proportionality constant $\rho_{0}$.
So the expected number of objects contained in volume $v$ is $\langle n(v) \rangle = \rho_{0} v$,
and $\rho_{0}$ that can be defined as the density of our pointlike objects is the unknown to be found.
An alternative treatment of our problem, which is the case of the nearest neighbors method, consists of finding the volume
corresponding to a predefined number of points $N$. The expected value of volume $\langle v \rangle$ is then proportional
to the number of points with a proportionality constant $\mu$ defined as the volume per point. So the expected volume
over $N$ points is $\langle v_{N} \rangle = \mu N$. In this case $\mu$ is the basic unknown.

Let us fix the origin of cartesian coordinates at an arbitrarily chosen point and imagine a series of $N$ hyperspheres
centered on the origin such that only one pointlike object resides on the surface of each hypersphere.
The sequence is ordered according to increasing volume $v_{1}, v_{2}, \ldots , v_{N}$.
It is easy to see that the problem is identical to the well known case of events occuring randomly but at a constant
average rate (e.g. Eadie et al. 1982). Examples often presented in statistical textbooks are radioactive decay,
telephone calls, or recording photons arriving from a faint astronomical object. In our case the volume plays
the role of time. It is worthwhile to point out that the variables in any pair of $v_{i}$ values are not statistically
independent because the central part of the larger hypersphere is identical with the smaller one. In order to deal with
statistically independent observables we will consider the first order differences of consecutive volumes

\begin{equation}
x_{i} = v_{i} - v_{i-1}   ~~~~~~~~    i = 1, 2, \ldots , N
\end{equation}
with $v_{o} = 0$. Random variables $x_{i}$ are mutually statistically
independent and their probability distribution is exponential. Therefore,
the corresponding probability density can be written as

\begin{equation}
                 f(x_{i}) = \frac{1}{\mu_{i}} e^{-\frac{x_{i}}{\mu_{i}}} .
\end{equation}

\noindent
For a uniform distribution of our pointlike objects all $\mu$ values are
identical and equal to $\mu_{0}$. Therefore the joint probability density
of all $x_{i}$ is

\begin{equation}
 L = f(x_{1}, x_{2}, \ldots , x_{N}) = \frac{1}{\mu_{0}^{N}} e^{-\frac{1}{\mu_{o}}\sum_{i=1}^{N}x_{i}} .
\end{equation}

\noindent
We can find a maximum likelihood estimator for the volume per point $\langle\mu_{0}\rangle$
\begin{equation}
              {\langle\mu_{0}\rangle} = \frac{\sum_{i=1}^{N}x_{i}}{N} =\frac{v_{N}}{N} .
\end{equation}

\noindent
This estimator is based solely on the position of the most distant neighbor in the sample.
Under the assumption of constant density, this estimator is sufficient and unbiased, i.e.
it already includes all information on density contained in our sample.
Exact positions of less distant neighbors are irrelevant.

The probability density of the random variable $v_{N}$ follows the Gamma distribution
\begin{equation}
    f(v_{N}) =\frac{v_{N}^{(N-1)}}{\mu_{o}^{N}\Gamma(N)} e^{-\frac{v_{N}}{\mu_{o}}}
\end{equation}
with expectation value $\langle v_{N} \rangle = \mu_{0} N$, variance $\sigma^{2}(v_{N}) = \mu_{0}^{2} N$,
and standard deviation $\sigma(v_{N}) = \mu_{0} \sqrt{N}$.
The corresponding values for the estimated value of $\mu$ are $\langle\mu\rangle = \mu_{0}$, variance $\sigma^2(\mu) = \mu_{0}^{2}/N$
and standard deviation $\sigma(\mu) = \mu_{0} / \sqrt{N}$.
It is worth noting that the variance of the estimated volume per point
$\mu$ is exactly equal the lowest possible value set by the sampling
statistics because it is inversely proportional to the number of independent 
observations $N$ and the variance is defined for all
values of $N$ starting with $N=1$.

The probability density of $v_{N}$ is given by Equation~5. We can treat all
the remaining volumes $v_{i}$ with $i<N$ as random variables uniformly distributed
between $0$ and $v_{N}$.
We can now write the joint probability density of volumes $v_{i}$ for uniform point distributions
\begin{equation}
L = f(v_{1}, v_{2}, \ldots , v_{N}) = f(v_{N})\prod_{i=1}^{N-1} f(v_{i} | v_{N}) = f(v_{N}) \frac{1}{(v_{N})^{N-1}} ,
\end{equation}
\noindent
where $1/v_{N}$ is the conditional probability density of any point other than $N$-th given that the volume $v_{N}$
is fixed.

Casertano and Hut (1985) derived analogous formulae for the alternative
case of point density estimation. They had to consider an inverse value
of directly observable $v_{N}$ that led to the use of the inverse Gamma
probability distribution instead of the Gamma distribution, and consequently to a loss
of information. The estimator of density is
\begin{equation}
   \langle\rho\rangle = \frac{N-1}{v_{N}}
\end{equation}
and the variance expressed in terms of the estimated density is
\begin{equation}
   \sigma^{2}(\rho) = \frac{\langle\rho\rangle^{2}}{N-2} .
\end{equation}
Therefore the above approach should only be used for $N>2$. The variance of this estimator
is larger than the sampling statistics limit, even drastically so
for very small $N$. Evaluating density at a location coinciding with one
of the pointlike objects is no different. Points at the origin of coordinates should
not be counted.

\Section{Non-uniform Density Distribution}

The optimal properties of the $v_{N}$ estimator degrade for density profiles with progressively larger deviations
from a uniform distribution. The smoothing bias is increasing. In addition, the random variables $x_{i}$ are no longer
mutually statistically independent and now the exact positions of less distant neighbors, normalized to the value of $v_{N}$,
carry information that can be used to limit the influence of smoothing bias.
An approximation formula based on power series expansion is a natural choice and coefficients can be estimated using
least squares. A less complicated approach is to use orthogonal functions constructed from the power series of the same order.
In the latter case the unknown coefficients can be determined by convolutions, which is both simpler and faster.

Let us consider the sequence of normalized volumes $y_i = v_{i}/v_{N}$ with the exception of the last element
taken as normalization. Spherical volumes in $D$ dimensions are computed as $v_i = r_i^D\pi^{D/2}/\Gamma(D/2+1)$,
where $r_i$ is the distance to the $i$-th nearest neighbor.
In the uniform case $v_{1}, v_{2}, \ldots , v_{N-1}$ are uniformly distributed
over the interval $(0, v_{N})$. Observed deviations from the uniform distribution can tell us something about
the density distribution inside the volume defined by the $N$-th point. We will search for this something
with the help of the Legendre polynomial expansion of order $k$. In the following derivation we make use
of the shifted Legendre polynomials $\tilde{P}_l(y) \equiv P_l(2y-1)$ which are orthogonal on the interval $(0, 1)$
and are obtained from the regular Legendre functions $P_l$ defined over the interval $[-1, 1]$.
The observables can be expressed with the help of the Dirac delta function
\begin{equation}
      \rho(y) = \frac{1}{v_{N}} \sum_{i=1}^{N-1}\delta (y-y_{i}) ,
\end{equation}
\noindent
and in this form are ready to be convolved with the shifted Legendre functions.
The first few basis functions are
\begin{eqnarray}
\tilde{P}_{0}(y) &=& 1\\
\tilde{P}_{1}(y) &=& 2y - 1\\
\tilde{P}_{2}(y) &=& 6y^2 - 6y + 1\\
\tilde{P}_{3}(y) &=& 20y^3 - 30y^2 + 12y - 1 .
\end{eqnarray}
A convolution with the $l$-th term yields the corresponding expansion coefficient
\begin{equation}
   \rho_{l} = (2l+1) \int_{0}^{1} \rho(y)\tilde{P}_{l}(y)dy = \frac{2l+1}{v_{N}}\sum_{i=1}^{N-1}\tilde{P}_{l}(y_{i})
\end{equation}
and the resulting interpolation formula for density is
\begin{equation}
   \rho(y) = \sum_{l=0}^{k} \rho_{l} \tilde{P}_{l}(y) .
\end{equation}
The extrapolated value of central density expressed in points per unit volume is
\begin{equation}
   \hat{\rho}_{N, k} = \rho(y)|_{y=0} = \frac{1}{v_{N}} \sum_{i=1}^{N-1} \sum_{l=0}^{k} (2l+1) \tilde{P}_{l}(y_{i}) \tilde{P}_l(0) .
\end{equation}
Using the fact that $\tilde{P}_{l}(0) = (-1)^l$ and returning to regular Legendre polynomials $P_l$ we obtain the final
formula for the estimator
\begin{equation}
   \hat{\rho}_{N, k} = \frac{1}{v_{N}} \sum_{i=1}^{N-1} \sum_{l=0}^{k} (-1)^l (2l+1) P_{l}(2y_{i}-1) ,
\end{equation}
which is convenient to evaluate numerically as the second sum is the value of the regular Legendre series of order $k$
with fixed coefficients taken at $2y_i-1$. For $k=0$ we recover the original $N$-th nearest neighbor estimator $\hat{\rho}_{N, 0} = (N-1)/v_N$.

The simple method presented above samples the orthogonal basis functions at $N-1$ points
ignoring their shape over the rest of the interval. We can imagine another approach
that utilizes all information in basis vectors, e.g. by integrating the Legendre polynomials
between pairs of consecutive data points. However, simulations performed using this alternative
method have not improved the final accuracy of density estimates.

\Section{Monte Carlo Experiments}

\begin{figure}[htb]
\centerline{\includegraphics[width=12.7cm]{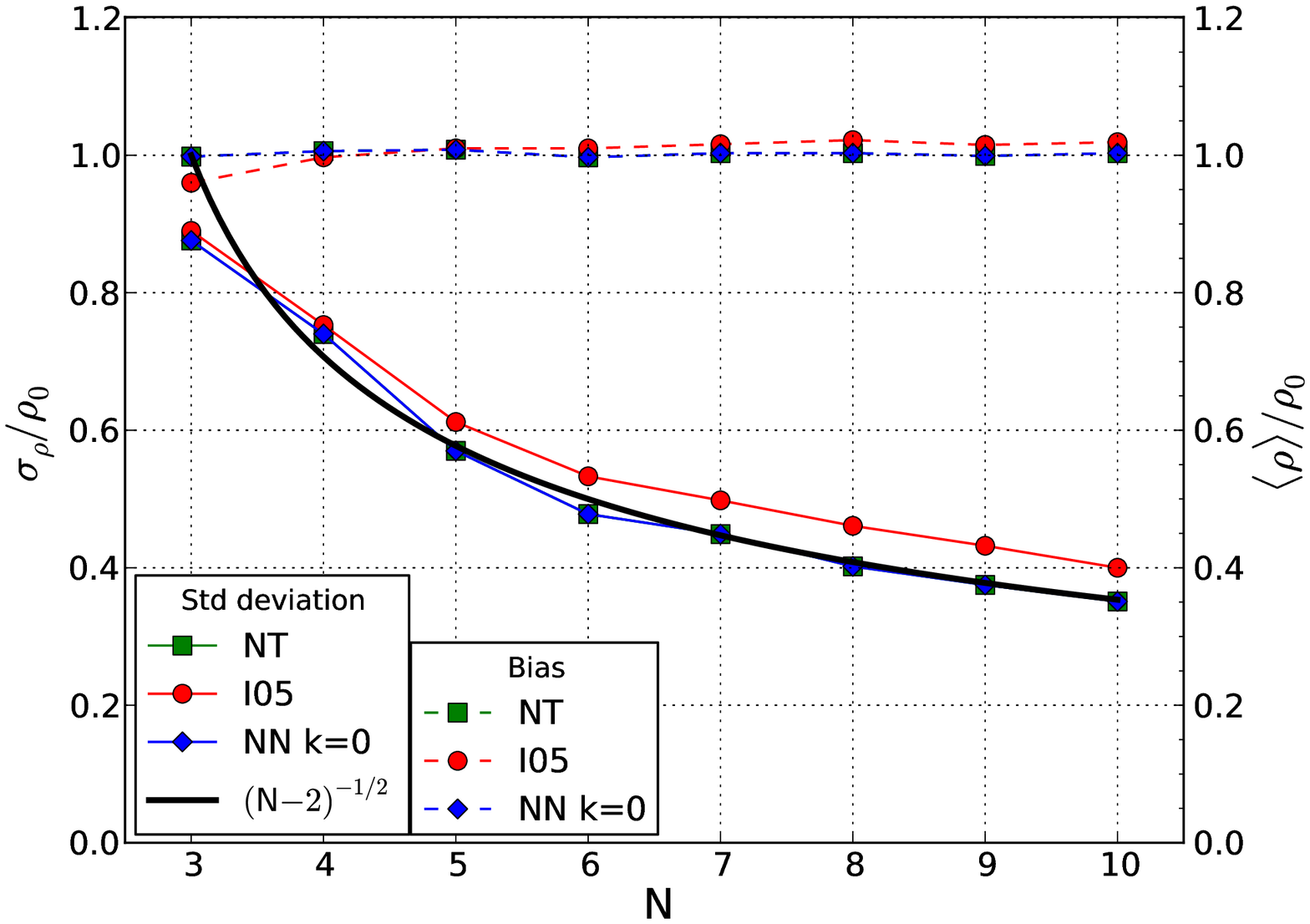}}
\vskip6pt
\FigCap{Performance of three density estimators based on the nearest neighbors method applied
to a uniform two-dimensional density distribution: 1) $N$-th nearest neighbor method (NT),
2) prescription of Ivezi\'c et al. (2005) (I05), and 3) our new algorithm based on interpolation
of $N$ nearest neighbors (NN) with density variations turned off ($k=0$). Standard deviation (solid lines)
and bias (dashed lines) were calculated as a function of $N$ by averaging the results of $10^4$ independent trials.
Theoretical noise limit of the inverse volume estimator is shown as the thick black curve.
}
\end{figure}

\begin{figure}[htb]
\centerline{\includegraphics[width=12.7cm]{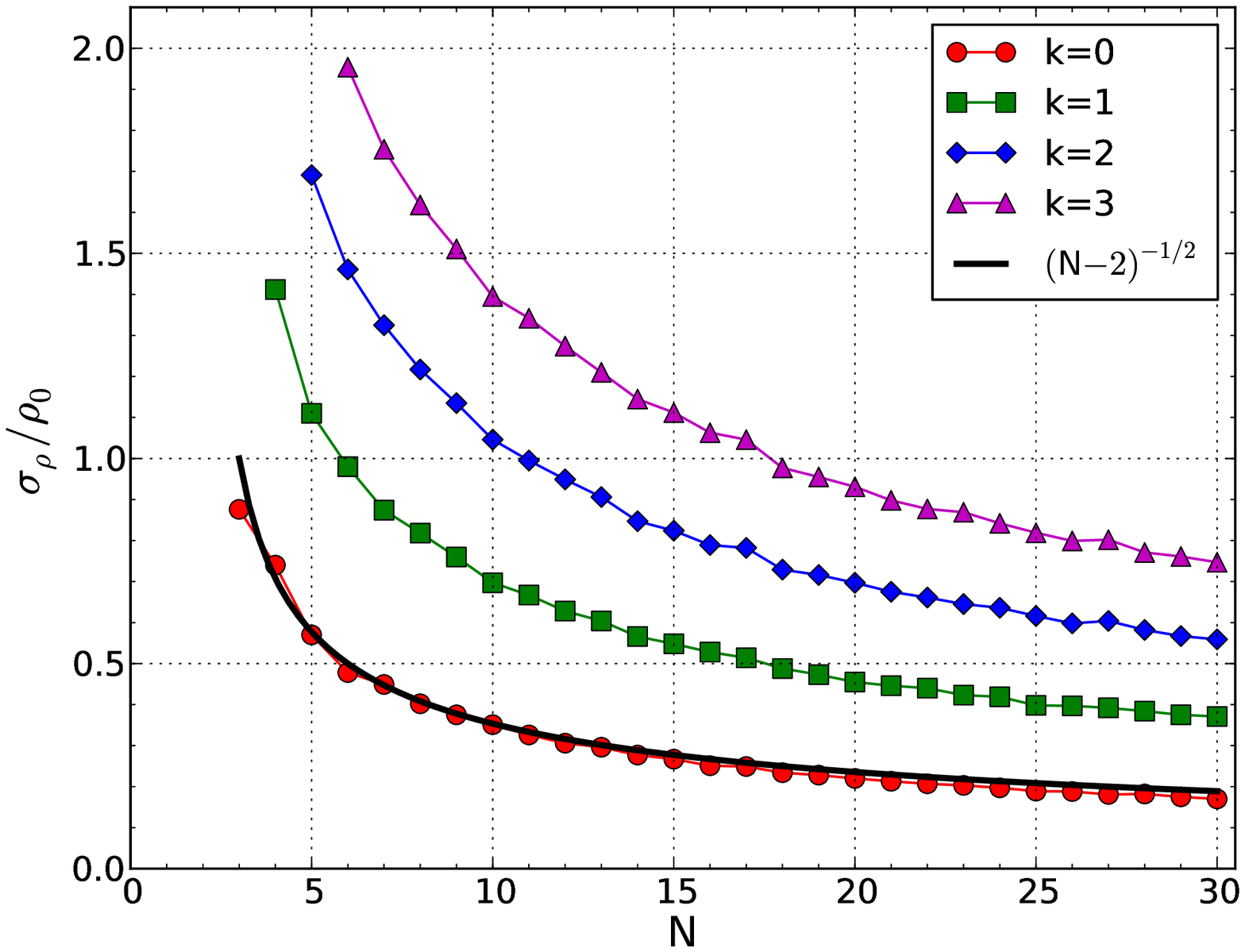}}
\vskip6pt
\FigCap{Accuracy of our modified density estimator based on the nearest neighbors method applied
to a uniform two-dimensional density distribution. Standard deviation was calculated as a function of the number
of nearest neighbors $N$ by averaging the results of $10^4$ independent trials.
The variance of the estimate degrades for methods of order $k>0$.
Theoretical noise limit of the inverse volume estimator is shown as the thick black curve.
}
\end{figure}

\begin{figure}[htb]
\centerline{\includegraphics[width=12.7cm]{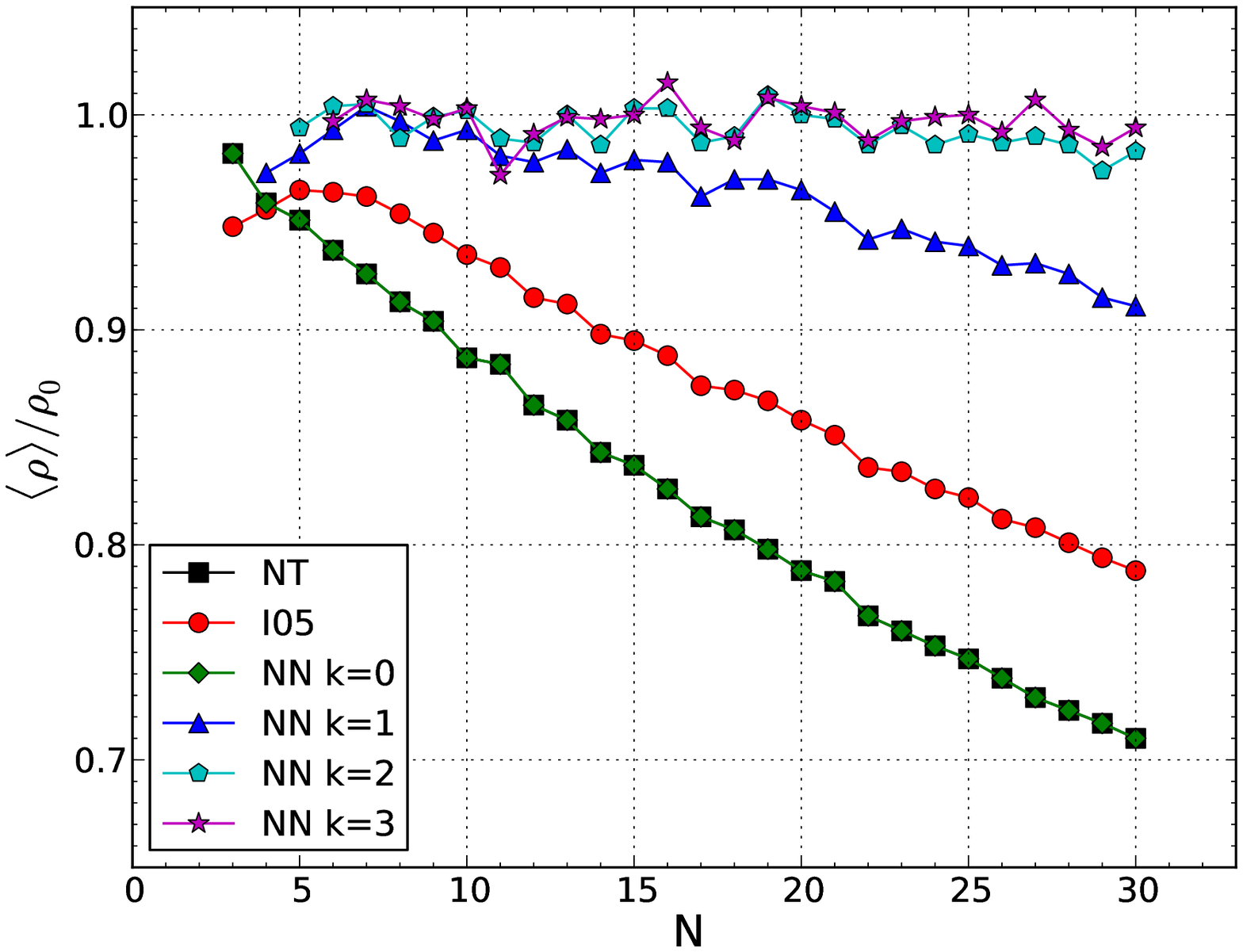}}
\vskip6pt
\FigCap{Comparison of the bias introduced by various density estimators based on the nearest neighbors method applied
to a non-uniform two-dimensional density distribution. The simulated density field consists of 10 points drawn from a Gaussian
distibution on top of a uniform background that doubles the density at the maximum. The lines show the ratio of the estimated density
and true density at the peak as a function of the number of nearest neighbors $N$, and were calculated by averaging the results
of $10^4$ independent trials.
}
\end{figure}

With the help of Monte Carlo simulations we can further investigate the properties of our new estimator.
In Figure~1 we compare the bias and standard deviation of three estimators: 1) $N$-th nearest neighbor
(hereafter NT), 2) Ivezi\'c et al. (2005) (I05), and 3) our $N$ nearest neighbors estimator from Section~3 (NN) with $k=0$.
The NN algorithm with $k=0$ is mathematically equivalent to NT estimator.
The experiment consists of $10^4$ realizations of a uniform two-dimensional density field. Each algorithm is applied
to the same data with $3 \leq N \leq 10$.
All three estimators are unbiased ($\langle n \rangle/n \simeq 1$) and follow the theoretical variance curve.
The performance of the I05 algorithm in this case is essentially the same as for the other two methods.
The standard deviation of higher order NN estimators is shown in Figure~2. The number of nearest neighbors $N$
varies between the lowest possible value ($k+3$) and 30.
Figure~3 demonstrates how the same three estimators handle non-uniform density distributions and smoothing bias.
On top of the uniform two-dimensional density field we now include a Gaussian peak that doubles the surface
density of points at the maximum. As before, we run $10^4$ density estimates at the location of the peak.
The overdensity at the center is sparsely sampled with only 10 data points drawn from the two-dimensional normal
distribution. The bias given as $\langle n \rangle/n$ is shown for $k+3 \leq N \leq 30$. As $N$ increases,
the estimators effectively average input data over larger areas. Again, the NN $k=0$ case is just the $N$-th nearest
neighbor algorithm.
The estimator of Ivezi\'c et al. (2005) can absorb some bias at the cost of increased variance.
Our new method is quite efficient in removing the smoothing bias and offers some flexibility in handling the tradeoff
between the bias and the variance of the estimator. The second order estimate is practically unbiased in this test.

\Section{Conclusions}

Using the nearest neighbors method we obtain $N$ independent observables $x_{i}$ with which to estimate
the density. We can treat them as $N$ information units as long as they are used to measure the volume
per data point. Inverting the observables and considering them as measures of density effectively lowers
the number of information units to $N-2$. Consequently, the variance of the estimator increases by $N/(N-2)$.
This excludes 1 and 2 as the allowed values of $N$. The information content of the observables $x_{i}$ cannot
be increased by transforming them.

In the non-uniform case the accuracy of our results is also affected by the smoothing bias,
in addition to the basic limitation due to the number of degrees of freedom.
The resulting density estimate is the average density inside the volume defined
by the most distant point in the sample and not the density at the chosen center.
The smoothing bias increases with increasing number of neighboring points $N$
supplied to the estimator. Therefore, the choice of $N$ is a tradeoff between the accuracy and
the smoothing bias.

In Section~3 we used the distribution of distances to $N$ nearest neighbors to ``fit''
a simple interpolation formula that captures local density variations around an arbitrary center.
A density estimate at the center is then obtained by extrapolating this formula to zero distance.
And it is this extrapolation that is responsible for a large increase in variance as higher order terms
are included in the density profile. However, increasing $k$ allows one to use larger $N$
while maintaining control over smoothing bias, which results in more accurate density estimates.
The best values of $N$ and $k$ for a particular application may be selected with the help of
Monte Carlo experiments.

\Acknow{This work was supported by the LDRD program at LANL.}

\end{document}